\documentclass[12pt]{article}

\usepackage{graphicx}
\usepackage{subfig}

\usepackage{placeins}

\usepackage{authblk}
\usepackage{amsmath}
\usepackage[margin=1in]{geometry}

\usepackage[math]{cellspace}
\newcommand{\tScale}{1}

\begin{document}

\title {Photons are lying about where they have been, again }

\author[1]{Gregory Reznik}

\affil[1]{\small Raymond and Beverly Sackler School of Physics and Astronomy, Tel-Aviv University, Tel-Aviv 69978, Israel}
\affil[2]{Max-Planck-Institut f\"{u}r Quantenoptik, Hans-Kopfermann-Stra{\ss}e 1, 85748 Garching, Germany}
\affil[3]{Department f\"{u}r Physik, Ludwig-Maximilians-Universit\"{a}t, 80797 M\"{u}nchen, Germany}

\author[2,3]{Carlotta Versmold}

\author[2,3]{Jan Dziewior}
\affil[4]{Schmid College of Science and Technology, Chapman University, Orange, California 92866, USA}
\affil[5]{Institute for Quantum Studies, Chapman University, Orange, California 92866, USA}

\author[2,3]{Florian Huber}

\author[1]{Shrobona Bagchi}

\author[2,3]{Harald Weinfurter}

\author[4,5]{Justin Dressel}

\author[1,2,3,5]{Lev Vaidman}

\date{}

\maketitle

\begin{abstract}
Bhati and Arvind [Phys. Lett. A, 127955 (2022)] recently argued that in a specially designed experiment the timing of photon detection events demonstrates photon presence at a location at which they are not present according to the weak value approach. 
The alleged contradiction is resolved by a subtle interference effect resulting in anomalous sensitivity of the signal imprinted on the postselected photons for the interaction at this location, similarly to the case of a nested Mach-Zehnder interferometer with a Dove prism [Quant. Stud.: Mat. Found. 2, 255 (2015)]. 
We perform an in depth analysis of the characterization of the presence of a pre- and postselected particle at a particular location based on information imprinted on the particle itself.
The theoretical results are tested by a computer simulation of the proposed experiment.

\end{abstract}

%\maketitle

%--------------------------------------------------------------------------
\section{Introduction}

%Standard quantum formalism usually avoids discussing quantum systems between measurements. The analysis of pre and postselected quantum systems is the objective of the two-state vector formalism (TSVF) \cite{ABL,AV90,AV91}.
The standard approach to quantum mechanics is concerned with probabilities of measurement outcomes and avoids discussing properties of quantum systems between measurements.
This is the subject of the two-state vector formalism (TSVF) \cite{ABL,AV90,AV91}, which provides a general analysis of the behavior of pre- and postselected quantum systems.
The TSVF was applied by Vaidman \cite{past} to consider the question of the location of a quantum particle in the past based on the weak trace analysis (WTA).  
Starting from the first experiment on this issue \cite{Danan} this analysis led to a considerable controversy. %, see e.g.~\cite{} for just a few examples.
%This analysis led to a considerable controversy, and in this paper we discuss the most recent argument against it. 

The most recent criticism has been brought forward by Bhati and Arvind (BA), who review the TSVF and propose an experiment designed to show an inconsistency in the WTA.
They introduce a different approach to quantifying presence based on information about the location imprinted on the travelling particle itself.
An application of this criterion to their proposed interferometric setup seems to yield a contradiction with the WTA concerning the presence of the pre- and postselected photons at one location in the setup.

Here, based on the scenario presented by BA, we develop an analysis of particle presence based on information imprinted on travelling particles. 
We agree with the predictions of the signal by BA in their experiment, but argue that in general the signal from a location is not a reliable indicator of a particle's presence at that location.
Since the signal is imprinted on the particle itself it might become distorted along the path of the particle, in particular, it can be amplified leading to a false indication of a strong presence.  
To check if such a distortion is present we propose to consider the signal imprinted on a particle fully localized at the location in question as a test case.
We introduce a method to quantify the amount of information in the signal and compare the information collected by a localized particle with the information imprinted on the pre- and postselected particle in question.
We argue that presence is indicated only if the amount of information carried by the particle in question is comparable to the amount of information carried by a localized particle.
It turns out that if the criterion based on imprinted information is analyzed properly in this manner, it agrees with the predictions given by the WTA. %, invalidating the criticism by BA.

The paper is organised as follows.
In Section \ref{sec::BAargument} we introduce the scenario and argument of BA.
In Section \ref{sec::photonslying} we compare it to the original experiment \cite{Danan} which demonstrated a surprising path of the photon and another scenario which claimed to demonstrate the inconsistency of the WTA \cite{Dove} and which turns out to be similar to the experiment of BA.
Then we demonstrate in three ways how the method of BA is anomalously sensitive to the presence at a particular location: a theoretical analysis of the predicted signals in Section \ref{sec::theo}, a numerical simulation of the experiment in Section \ref{sec::fit}, and in Section \ref{sec::simplified} a simplified scenario which allows a more direct procedure of information extraction leading to a clearer analysis of the information content.
Section \ref{sec::summary} summarizes our results.

%OLD--------------------------------------------------------------------------

%We agree with the BA calculations of the signal in their experiment, but argue that in general the signal from a location is not a reliable indicator of a particle's presence at that location because it might become amplified along the path of the particle, leading to a false indication of a strong presence.  
%We claim that proper quantification of the  presence based on signatures imprinted on the particle requires introducing of a reference benchmark of the information content of a quantum particle. 

%The paper is organised as follows. In Section \ref{sec::BAargument} we introduce the scenario and argument of BA.
%In Section \ref{sec::photonslying} we compare it to the original experiment \cite{Danan} which demonstrated a surprising path of the photon in the past and another scenario which claimed to demonstrate the inconsistency of the WTA \cite{Dove} and turns out to be similar to the BA example.
%Then we demonstrate in three ways how the method of BA is especially sensitive to the presence at a particular location: a theoretical analysis of the predicted signals in Section \ref{sec::theo}, a numerical simulation of the experiment  in Section \ref{sec::fit}, and in Section \ref{sec::simplified}  a simplified scenario which allows a more transparent analysis. 
% Section \ref{sec::summary} summarizes our results.

%--------------------------------------------------------------------------
%SUMMARY OF BA ARGUMENTS
\section{Argument by Bhati and Arvind}
\label{sec::BAargument}
BA define three statements:
\begin{itemize}
\item[S-A:] If the weak value of the projection operator $\Pi _x =|x\rangle\langle x|$ at an intermediate time is zero, where $|x\rangle$ is a position eigenstate, then the particle was not present at position $x$ at that time.

\item[S-B:] A quantum particle was present at a location if and only if it left a weak trace on a pointer located at that location upon interaction.

\item[S-C:] A quantum particle cannot carry information about a localized object without interacting with it.
In particular, if the particle is a photon inside an interferometer, it cannot not visit the location of a localized optical device and still gain information about it.
\end{itemize}

Statements similar to S-A and S-B are indeed the fundamental elements of the WTA, where S-B is an operational criterion defining particle presence and S-A relates the TSVF formalism to the concept of presence.
%One possibility to state the WTA approach is the equivalence of these two definitions of presence.
S-C represents another operational definition of presence which, in contrast to S-B, does not refer to traces on the local environment. S-C, instead, considers the traces left by the local interaction  on the travelling particle itself.
(Using internal particle degrees of freedom as pointers is actually how the majority of weak value experiments has been performed in the past, see e.g.~\cite{KwiatSpinHall,
Dixon,PNAS}
with \cite{Groen,Steinberg} being notable exceptions.)
%Using a conception of presence along the lines of S-C is actually how the majority of weak value experiments has attempted to address the WTA in the past \cite{KwiatSpinHall,BoydAngRot,Danan,PNAS,Sequential}, with \cite{Steinberg} being a notable exception.

BA claim that in their (gedanken) experiment, see Fig. \ref{fig::BAfig1}a, the following happens:
\begin{itemize}
\item [i)] The postselected photons carry information about the frequency of the modulation at location $L_1$.

\item [ii)] The weak value of projection of every photon on $L_1$ is vanishingly small.

\item [iii)] The weak local trace at $L_1$ is  vanishingly small.
\end{itemize}

They argue that from S-C and (i) follows that the photons were at $L_1$.
Conversely, from S-A and (ii), as well as from S-B and (iii), it follows that the photons were not at $L_1$.
The contradiction between S-C and S-A, as well as between S-C and S-B puts in question the WTA of the past of a quantum particles \cite{past}.

%--------------------------------------------------------------------------
%ARGUMENT 1 - Complications when using particle DOF as pointer

\section{Photons lying about where they have been}
\label{sec::photonslying}

The origin of the confusion is the title of the experimental paper ``Asking photons where they have been'' \cite{Danan} demonstrating the theoretical results of \cite{past} regarding the traces the photons leave in a nested Mach-Zehnder interferometer.
Due to the general experimental difficulty of observing the local trace pre- and postselected quantum particles leave on external systems, the local trace was demonstrated, instead, via an observation of the trace left on the photon itself.
The justification for this indirect method is that the change of the photon's degrees of freedom was created locally at the location in question and this degree of freedom was not further disturbed until the measurement.
%was transmitted without disturbance to the location of the measurement.
This process, however, happened to have also another interpretation, which (maybe unfortunately) was chosen for the title of \cite{Danan}, the interpretation described by S-C.
%Due to experimental difficulties of observing the local trace a quantum particle leaves on external systems, the locally created particular trace on the degrees of freedom of the particle itself as referred to in S-C was observed instead.

\begin{figure}
\centerline{\includegraphics[width=1\textwidth]{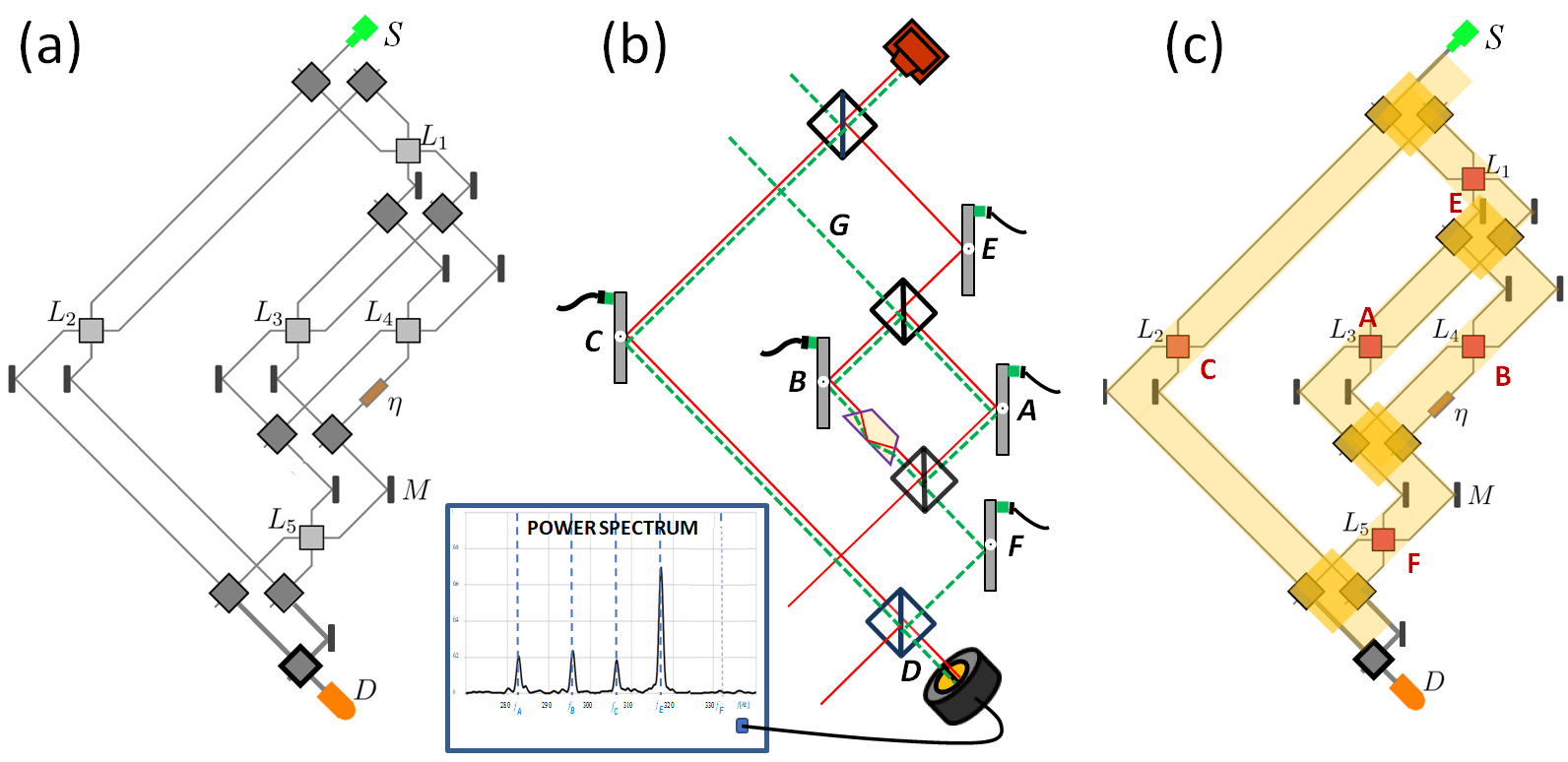}}
\caption{ {\bf The BA and Dove prism interferometers.}
(a) The BA interferometer (based on Fig.~1 of \cite{Bhati}).
Six-port interferometer with single photons sent from source $S$ and detected in detector $D$.
The dark square boxes are the beamsplitters, the light boxes ($L_i$) are the time dependent beamsplitters with transmission coefficients $\sin(\epsilon\cos \omega_i t)$ and the long dark rectangle are the mirrors.
(b) Nested Mach–Zehnder interferometer with a Dove prism in arm $B$ (based on Fig. 2 of \cite{Lying}). 
The region of the overlap of the forward (continuous red) and the backward (dashed green) evolving states is where the photons are present according to WTA, but the predicted results of the experiment include a strong signal from mirror $E$, where detected photons were not supposed to be.
(c) Analogy between the BA and the Dove prism interferometers.
The thick yellow lines represent arms of the Dove prism interferometer with two modes: the undisturbed Gaussian mode and the orthogonal mode.
These modes correspond to the pair of channels of the BA interferometer.
The arms A, B, C, E, F of the Dove prism interferometer correspond to $L_3$, $L_4$, $L_2$, $L_1$, $L_5$ of the BA interferometer.
The phase shifter $\eta$ in BA setup corresponds to the Dove prism and shifts the phase by $\pi$.
The detector of the BA interferometer corresponds to the upper part of the dual cell detector in the Dove prism experiment.
} 
\label{fig::BAfig1}
\end{figure}

Alonso and Jordan \cite{Dove} were first to present an example in which S-C apparently contradicts the WTA.
They pointed out that introducing a Dove prism in the experiment \cite{Danan} does not change the WTA description of where the photons have been, but leads to different experimental results: the photons provide information about a location where they have not been according to the WTA.
Vaidman and Tsutsui (VT) \cite{Lying} explained that introducing the Dove prism in the setup of \cite{Danan} makes the photons ``lie about where they have been''.
In the experiment \cite{Danan} the locally created trace on the transversal motion degree of freedom of photons was not distorted during the evolution from the location of the interaction until the detection.
This justified presenting the signal from the photons as the local weak trace.
Introducing the Dove prism, see Fig.~\ref{fig::BAfig1}b, spoils the experiment as an observation of the local trace because the transversal degree of freedom of the photons is distorted on their way from $E$ to the detector.
VT argue that in fact in the experiment with the Dove prism the signal carrying information about $E$ appears similar (or even bigger) than the signals from the locations $A$, $B$, and $C$, where the photons have been, because of an anomalous sensitivity of the pre- and postselected photons to the interaction at $E$. %{\color{red} Intuitively, the probe disturbs the configuration so that a few photons do leak into the region $E$, where they receive an \emph{amplified} perturbation; thus, the total signal from $E$ can appear misleadingly large even though it is still being greatly suppressed relative to the signal one would have gotten if the number of photons traversing $E$ had actually been the same as at $A$, $B$, or $C$. The collected photons are essentially ``lying'' about how many could be found at each location.} 
We argue that the same is the case in the experiment proposed by BA.

%--------------------------------------------------------------------------
%ARGUMENT 2 - Extreme sensitivity as complication to be considered => i) is wrong and thus S-C does not conclude presence

%2.1 Correspondence of the experiments
The similarity between the experiment with the Dove prism, Fig.~\ref{fig::BAfig1}b, and BA experiment, Fig.~\ref{fig::BAfig1}a, is demonstrated in  Fig.~\ref{fig::BAfig1}c.
VT considered two orthogonal transversal spatial modes $\chi$ and $\chi_\perp$  in every arm of the interferometer with the Dove prism.
The six degrees of freedom of BA correspond to these two orthogonal modes in each of the arms $C$, $A$, and $B$.
The location $L_2$ corresponds to $C$, $L_3$ corresponds to $A$, and $L_4$ corresponds to $B$.
Furthermore, location $L_1$ corresponds to $E$ and $L_5$ corresponds to $F$, see Fig.~\ref{fig::BAfig1}c.
The role of the Dove prism is played by the phase shifter $\eta$.
In the VT experiment the signal is the difference between counts in the upper and lower parts of the spatial detector which correspond to different output ports  of the BA experiment.
In the BA experiment, instead, the detector is placed only in one output port.
This change in postselection slightly changes the situation, but the essential features of the experiments remain the same.
In the BA experiment, as in the VT case, the signal from $L_1$ is of the order of the signals from $L_2$, $L_3$, and $L_4$  where the photons have been, not because the photons were at $L_1$, but because the setup is much more sensitive to the interaction at $L_1$.
The photons ``lied'' that they were at $L_1$. 
All statements, S-A, S-B and S-C, are oversimplified and need clarifications, but it is S-C which leads BA to the wrong conclusion.% {\color{red} since it does not adequately account for the role of the probe disturbance or the possible amplification of its interaction strength}. 

%2.2 Critique and Clarification of Statements S-A, S-B and S-C => S-C is where they draw a wrong conclusion

\section{Theoretical analysis of Bhati-Arvind experiment}
\label{sec::theo}

\subsection{Criteria for presence of pre- and postselcted photons}
\label{sec::theo::criteria}
In the formal criterion S-A the condition of the weak value being 0 is fulfilled only in the case of zero interaction.
BA themselves provide a nonvanishing expression for the weak value in the first row of their Eq.~(10).
The value is of first order in $\epsilon$, which they correctly argue can be disregarded.
Note also that S-A, as stated, is applicable only when other degrees of freedom are not involved.
In the general case, the projection operator $\mathbf{P}_x (= \Pi_x)$ in S-A should be replaced by any local operator $\mathbf{O} \mathbf{P}_x$, see Sec. VI of \cite{past}.
In fact, the projection which BA consider, $\Pi_3^w (t_2)$, is not a projection on $L_1$, but on one channel in $L_1$.
However, in their experiment the weak value of any local operator $\mathbf{O} \mathbf{P}_{L_1}$ is indeed not more than of first order in $\epsilon$, so, according to the WTA, the particle was not present at $L_1$.

When formulating the operational criterion S-B it is crucial to specify the magnitude of the weak trace.
The weak trace manifests the presence of the particle if and only if it is of the same order as the trace that a well localised particle at this location would leave \cite{past}.
The weak value of local variables at $L_1$ is of order $\epsilon$, and since the weak coupling considered by BA is also of order $\epsilon$, the resulting weak trace is of order $\epsilon^2$, much smaller than the weak trace of a localised particle at $L_1$, which is of order $\epsilon$.
Again, the particle is not present at $L_1$ according to the WTA, confirming the consistency of S-A and S-B.

BA claim that contrary to S-A and S-B the criterion S-C seems to indicate presence at $L_1$ since the frequency $\omega_1$ is present in the signal.
Indeed, if S-C is understood as a binary criterion, i.e.~a particle can either be present or not, then S-C is in contradiction with the other two.
There are general arguments against a binary concept of presence, e.g.~it would assign full presence to locations where only a vanishing tail of a quantum wave of a particle was present.
Since in most formal descriptions wave functions of quantum particles are extended to infinity this would make it impossible to consider particle presence at all.
In fact, in their paper BA themselves perform a quantitative analysis of the information carried by photons, which goes beyond a binary approach.
So, the question is not just whether any information about a particular location is carried by the particle, but crucially about the amount of carried information. %amount of information matters
The information might get changed along the further evolution of the particle, and this has to be taken into account when quantifying the amount of information at detection. %to understand amount of information we need reference
To this end we propose to consider the information carried by a particle in the same setup which is additionally conditioned to be fully localized at $L_1$.
Then, it is natural to consider the particle to be present at $L_1$ if it gained information about $L_1$ comparable to the information gained by a particle well localized at this place. %what is the best reference
Note, that even if the particle has no presence at a certain location according to this definition, it still might have ``secondary presence'' \cite{tracingthepast} there, gaining information about this location in lower order.

There is no generally accepted measure for the amount of information carried by a photon in BA type experiment.
%A single photon carries usually very little information which make it difficult to consider.
%An experiment with a large number of identical pre- and postselected photons provides reliable (but usually never certain) information about a local parameter.
In such experiments we obtain information not from a single photon but from an ensemble.
A large number of identical pre- and postselected photons provides reliable (but usually never certain) information about a local parameter.
We define the amount of information $\mathcal{I}$ carried by each photon in that experiment as
\begin{align}\label{infodefi}
\mathcal{I} \equiv {1}/{N_\mathrm{min}}
\end{align}
where $N_\mathrm{min}$ is the minimal number of pre- and postselected photons required to obtain local information with a predefined precision. % using an optimal strategy.
%a local property with a sufficiently high precision with an error less than...
%the information about a local property
%the local parameter
%with an error less than a small threshold value we choose to be $1\%$.
This is a somewhat arbitrary definition which is not easy to apply, mostly due to the difficulty of finding the optimal strategy of information extraction from the photon. %for any particular scenario
However, for our comparative analysis of different situations there are natural efficient strategies for extracting information, so our definition provides a sufficiently good estimate of the amount of information carried by photons in the discussed experiments. %especially if the estimated difference in information content is sufficiently large.
We will show that the information about the disturbance at $L_1$ carried by photons in the experiment of BA is significantly smaller than for photons fully localized at $L_1$ and disturbed in the same way.
%First, we provide a theoretical argument and then, in Section \ref{sec::fit},  we demonstrate this based on a procedure which extracts the information via fits to data obtained from a simulation of the BA experiment.
% Finally in Section \ref{sec::simplified} we show the same relationship again, this time based on a  version of the BA experiment with a much simpler encoding of the local parameters in the signal, allowing for a more direct procedure of information extraction.

\subsection{Comparison of the information carried by the photons in Bhati-Arvind experiment and by localized photons}
%2.3 Operational Definition of Sensitivity (and Prescription for Testing it)
The method of BA for obtaining information is measuring the number of photons detected by $D$ at different time windows.
They derive the probability of detection of the photon sent at time $t$ (their Eq.~(8)) as
\begin{align}\label{BA}
    P^{BA}\approx  \frac{1}{18}[1+2\epsilon (2\cos \omega_1t-\cos \omega_2t+\cos \omega_3t+\cos \omega_4t)].
\end{align}
The external parameter present only at $L_1$ is the frequency of the modulation $\omega_1$, so the amount of information about $\omega_1$ carried by the postselected photon can quantify its presence at $L_1$.
The amount of the information should be compared with the case of a photon fully localised at $L_1$ which undergoes the same local interaction.
There are two channels at $L_1$, so presence at $L_1$ does not specify fully the state of the photon. To ensure the state of the photon at $L_1$ to be as it is in the BA experiment
 we consider the BA setup with the addition of a nondemolition measurement of the presence of the particle in $L_1$ and take into account only the cases in which, in addition to the pre- and postselection of the original protocol (input in port $S$ and detection by $D$), the nondemolition measurement finds the particle at $L_1$.
 Nondemolition measurement means here a standard von Neumann measurement of the projection operator of the photon on the location $L_1$. If the photon is in an eigenstate of this operator, we will know the eigenvalue and in both cases the quantum state of  the photon will not be changed. As the photon is not destroyed, in a recent implementation of such a measurement \cite{Rempe} it was named ``nondestructive measurement''.
In this case the probability of detection in $D$, conditional on successful localisation at $L_1$, is
\begin{align}\label{non_L1}
    P ^{L_1}\approx \frac{1}{12}\epsilon^2(2\cos (\omega_1t)+\cos (\omega_3t)+\cos (\omega_4t))^2.
\end{align}

%2.4 Formal argument for hyper sensitivity
To apply condition S-C, the quantity of interest is the amount of information carried by the pre- and postselected photons, i.e.~the information content per detected photon.
%Thus, it does not matter how many photons we have to send to obtain the postselected photon.
%the total number of successfully postselected particles is of no importance.
In the case of the BA experiment, the expression (\ref{BA}) for the probability of detection in $D$ includes a constant term which is not present in the expression (\ref{non_L1}) for photons localized at $L_1$.
This constant term carries no information about any of the frequencies $\omega_i$ and is larger than the information carrier terms by a significant factor of order $1/\epsilon$.
Therefore, we should expect that we will need to detect many more pre- and postselected photons in the original BA setup to gain the same information about $\omega_1$ than in the modified BA setup with localized photons.
This implies that the amount of information about $\omega_1$ per postselected photon is clearly smaller in the unmodified BA setup.
Note, that a small overall factor in the probability of postselection (\ref{non_L1}), which is independent of the parameters $\omega_i$, does not matter when quantifying information for postselected photons. %since only postselected photons are considered, 
%; i.e.,~ {\color{red} on average} each photon carries far less information about the process at $L_1$ in the BA experiment than the benchmark provided by photons fully localized at $L_1$. 
%{\color{red} Intuitively, most photons that reach the detector did not actually traverse $L_1$, as predicted by the TSVF, so do not provide any information about $\omega_1$; only the small fraction that did leak into $L_1$ due to the disturbance of the probe carry information about $\omega_1$.}

Disregarding the difference in the number of postselected photons between their experiment and the experiment with photons localized at $L_1$ is apparently the main reason for the mistaken conclusion of BA. 
Performing that comparison correctly shows that the presence of photons at $L_1$ is indeed being suppressed as claimed.
It seems that they also missed a statistical fluctuation term of the order of $\sqrt{N_s}$ in Eq. (9) (where $N_s$ is the number of incoming particles per time window), so their requirement to see the signal, $\epsilon N_s \gg 1$, has to be modified to $\epsilon \sqrt{N_s} \gg 1$.
However, even if this requirement is not fulfilled, we can estimate $\omega_1$ by increasing the duration of the experiment.

\subsection{Anomalous presence of a pre- and postselected particle}
\label{sec::anomalouspresence}

We want to mention an apparent inconsistency of our argument in the case of photons localized at $L_1$ described by Eq.~(\ref{non_L1}).
Our explanation was that in the BA experiment the sensitivity for the interaction at $L_1$ is much higher than for interactions at $L_3$ and $L_4$ but the formula shows that the signal regarding these locations is of the same strength, if the particle is localized at $L_1$.
The  reason for this is, however, that in this case the presence of the photon at $L_1$ is 1, while it is much larger at $L_3$ and $L_4$, since the weak values of projections on the arms of a two-path interferometer are strongly amplified for photons detected at the dark port, see \cite{PNAS}.
The presence at $L_5$ is also 1 but the sensitivity is not  increased and thus the signal with information about $\omega_5$ appears only in the next order of $\epsilon$.
It is contained in the next term omitted in (\ref{non_L1})
\begin{align}\label{addterm}
    \frac{\epsilon^3}{12}(2\cos\omega_1t+\cos\omega_3t+\cos\omega_4t)^2(\cos\omega_3t-\cos\omega_4t+2\cos\omega_5t).
\end{align}

Both in the BA experiment and in the experiment with localization at $L_1$ there is an anomalous sensitivity to the disturbance at $L_1$ and not to the disturbance at $L_4$.
Nevertheless, in each of the experiments comparable amounts of information about the locations $L_1$ and $L_4$ are carried by photons.
The explanations for these similar signals, however, are different.
In the BA case, the increased sensitivity at $L_1$ is countered by the tiny presence at $L_1$, while for photons localised at $L_1$, it is balanced by the anomalously large presence at $L_4$.
This anomalous presence at $L_4$ does not occur in the BA experiment without localization at $L_1$.

\section{Numerical simulation of Bhati-Arvind experiment}
\label{sec::fit}
\subsection{Simulation procedure}

While the qualitative  analysis above strongly implies our conclusion about the significant difference in the amount of information per particle, it is non-trivial to derive an exact analytical expression for the difference due to the different forms of the two probability functions (\ref{BA}) and (\ref{non_L1}).
In the following we demonstrate this difference by performing a computer simulation of the experiment and applying a reasonable method which extracts information about $\omega_1$ from simulated experimental data.

\begin{figure}
\centering
\subfloat{
    \includegraphics[width=\textwidth]{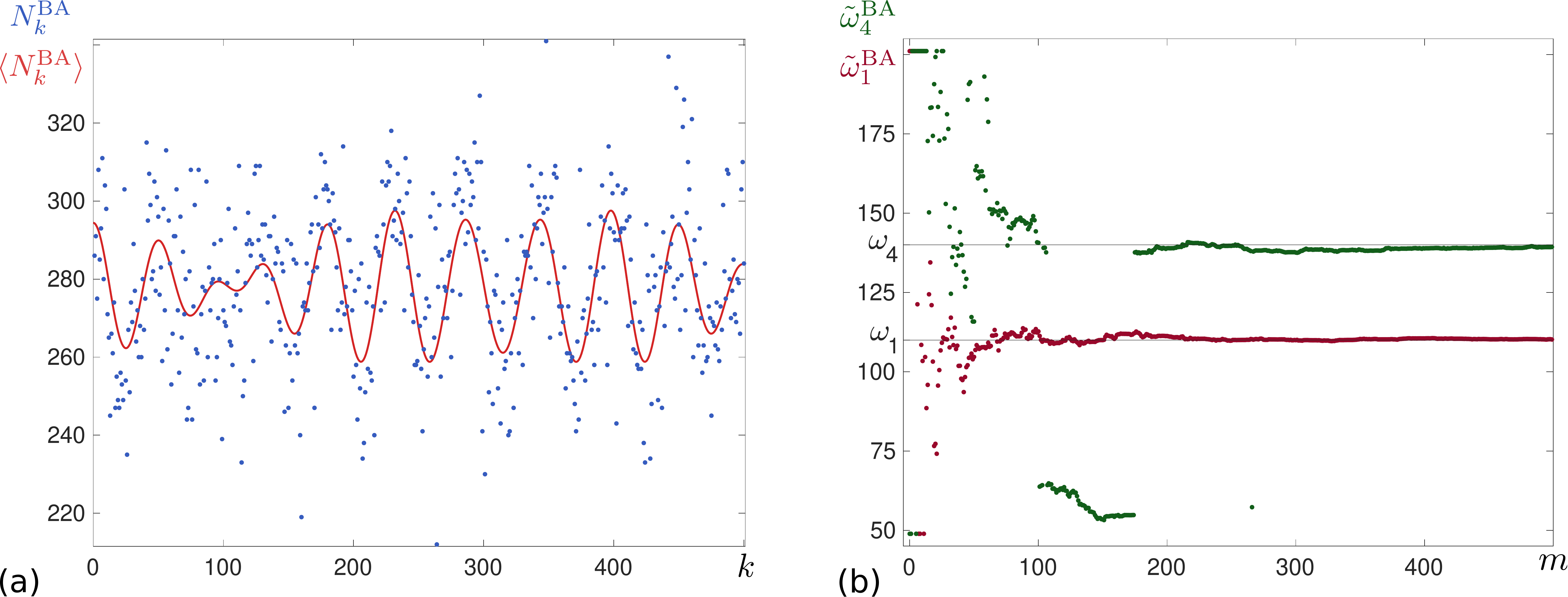}}
\caption{ {\bf Estimation of the frequency of the modulation in BA experiment.}
(a) The points provide the numbers of detected photons $N_k^{BA}$ in every time window $k$ generated 
by a computer simulation (implemented in MATLAB) of sending $N_s$ photons with detection probability (\ref{P_k}).
The continuous line is the theoretical expectation value $\langle{N}_k^{BA}\rangle$.
(b) The datapoints represent the results of the estimation algorithm for the frequencies based on the simulation data if only the first $m$ time windows are taken into account.
The red points show the case of estimation of $\omega_1$ (with all other parameters known), and the green points the estimation of $\omega_4$ (with all other parameters known).
The two horizontal lines mark the actual frequencies $\omega_1$ and $\omega_4$ employed in the simulation. Much better convergence of the estimation of $\omega_1$ than estimation of $\omega_4$ is expected due to the factor 2 between terms $\cos \omega_1 t$ and $\cos \omega_4 t$ in (\ref{BA}). } 
\label{fig:BA}
\end{figure}

%2.5 Simulation to show hyper sensitivity of L_1
Following BA we approximate the probability $P_k^{BA}$ for a photon sent in time window $k$ to reach the detector $D$ by the expression
\begin{align}\label{P_k}
    P_k^{BA} &= \frac{1}{18}+\frac{\epsilon}{9}\Big[ 2\cos \{\omega_1 (k-\frac{1}{2}) T_s\} -\cos \{\omega_2(k-\frac{1}{2}) T_s\} +\cos \{\omega_3(k-\frac{1}{2}) T_s\}+\cos \{\omega_4(k-\frac{1}{2}) T_s\}\Big],
\end{align}
where we choose interaction strength $\epsilon$, timestep $T_s$, frequencies $\omega_i$, and number of preselected photons per timestep $N_s$ as
\begin{align}\label{parameters}
    \epsilon=10^{-2}, \quad \quad T_s=10^{-3}\,{\rm s}, \quad \quad \omega_i=(100+10i)\,{\rm s}^{-1}, \quad \quad N_s=5000.
\end{align}
According to our understanding this choice of parameters fits the BA proposal.
The computer uses a random generator to simulate sending $N_s$ photons for each time window $k=1, 2,...500$ creating a series of numbers of postselected photons $N_k^{BA}$.
The results are presented in Fig.~\ref{fig:BA}a together with the theoretical expectation value $\langle {N}_k^{BA} \rangle = N_s P_k^{BA}$. 

%from the first through the $m$\textsuperscript{th} time window
The probability to obtain a particular series of postselected photon numbers for the first $m$ time windows, $\{N_1^{BA}, N_2^{BA},...,N_m^{BA}\}$, is
 \begin{align}\label{probability}
    \mathrm{prob}\Big[\{N_1^{BA}, N_2^{BA},...,N_m^{BA}\}\Big] = \prod_{k=1}^m \binom{N_s}{N_k^{BA}} (P_k^{BA})^{N_k^{BA}} (1-P_k^{BA})^{N_s-N_k^{BA}}~.
     %\mathrm{prob}\Big(\{N_1^{BA}, N_2^{BA},...,N_m^{BA}\}\Big) = \prod_{k=0}^m \binom{N_s}{N_k^{BA}} (P_k^{BA})^{N_k^{BA}} (1-P_k^{BA})^{N_s-N_k^{BA}}
\end{align}
%about $\omega_1$ 
To evaluate the information which is gained from the dataset $\{N_1^{BA}, N_2^{BA},...,N_m^{BA}\}$, we 
vary the parameter $\omega_1$ to
numerically maximize the probability (\ref{probability}) in order to obtain an estimate $\tilde{\omega}^{BA}_1$ and compare it to the actual value ${\omega}_1$.
%{\color{red} That is,} we maximise in (\ref{probability}), the probability to obtain {\color{red} the observed sequence} $\{N_1^{BA}, N_2^{BA},...,N_m^{BA}\}${\color{red}, while varying the estimate $\tilde{\omega}_1$ as the only free parameter after the substitution of (\ref{P_k}).}
We use a constrained optimization with $\tilde{\omega}^{BA}_1\in[48,202]\, \rm{s}^{-1}$.
The actual frequencies are far from the boundaries of this region, so our constraint does not affect the procedure significantly.
%$\tilde{\omega}^{BA}_1$ in the interval
%(The dependence of (\ref{probability}) on $\omega_1$ enters through substitution of (\ref{P_k}).)
This optimization procedure is repeated for each $m$ to produce the sequence of estimated values $\tilde{\omega}_1^{BA}(m)$ shown in Fig.~\ref{fig:BA}b in red dots.

\begin{figure}
\centering
\subfloat{
\includegraphics[width=\textwidth]{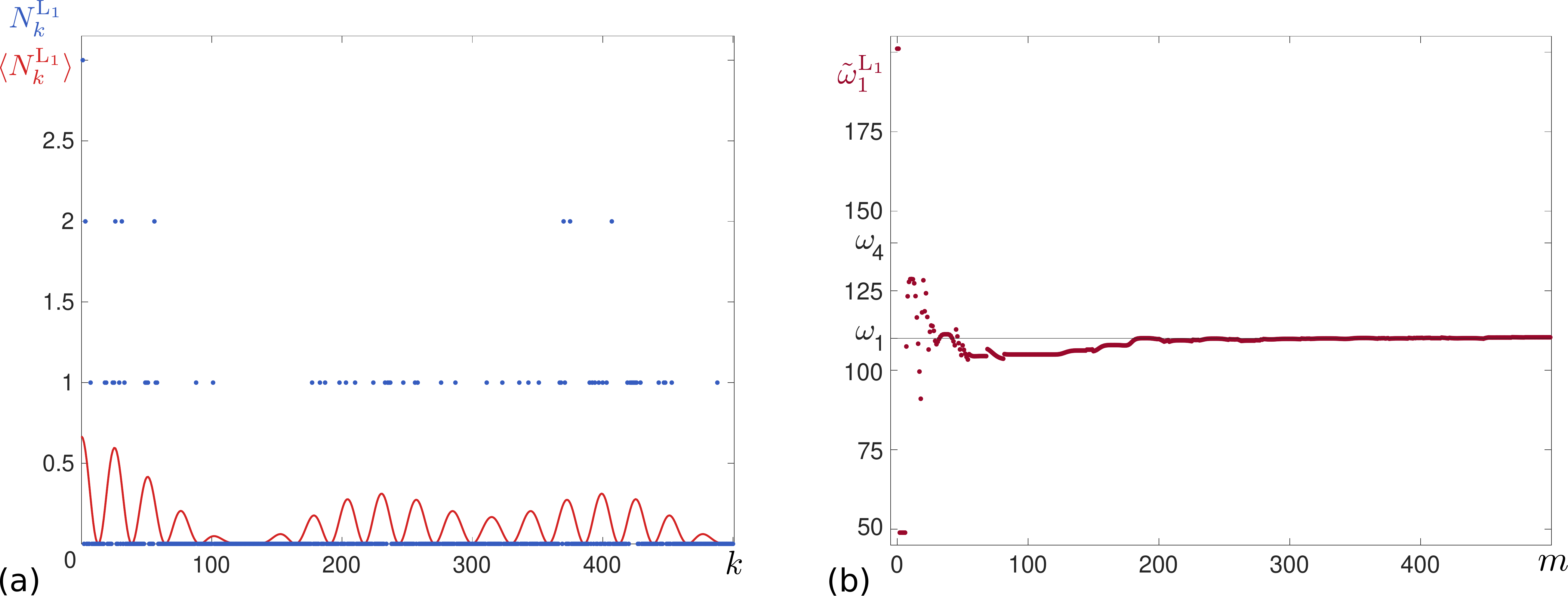}}
\caption{ {\bf Estimation of the frequency of the modulation at $L_1$ based on photons passing through $L_1$.} 
(a) 
The points provide the numbers of detected photons $N_k^{L_1}$ in every time window $k$ conditioned on their localisation at $L_1$ generated 
by a computer simulation  of sending $N_s$ photons with detection probability (\ref{P_kL_1}).
The continuous line is the theoretical expectation value $\langle{N}_k^{L_1}\rangle$.
(b) The datapoints represent the results of the estimation algorithm for the frequency $\omega_1$ (with all other parameters known).}
\label{fig:ND1}
\end{figure}

To obtain a benchmark sensitivity for comparison we repeat the simulation with the same parameters for the particles which have been localized at $L_1$.
The basis for this procedure is the probability function (\ref{non_L1}) which yields the probability distribution
\begin{align}\label{P_kL_1}
     P_k^{L_1} &= \frac{\epsilon^2}{12}\Big[ 2\cos \{\omega_1 (k-\frac{1}{2}) T_s\} +\cos \{\omega_3(k-\frac{1}{2}) T_s\} +\cos  \{\omega_4(k-\frac{1}{2}) T_s\}\Big] ^2.
\end{align}
The corresponding dataset $ \{N_1^{L_1}, N_2^{L_1},...,N_{500}^{L_1}\}$ together with the theoretical mean $\langle {N}_k^{L_1} \rangle = N_s P_k^{L_1}$ are presented in Fig.~\ref{fig:ND1}a. 
The estimated values $\tilde{\omega}_1^{L_1}(m)$ are shown in Fig. \ref{fig:ND1}b.

\subsection{Hyper-sensitivity at location $L_1$}
\label{ssec::hyperSensL1}

Since the frequencies marking the various locations in the experiment differ in steps of $10s^{-1}$ we obtain sufficient information about the corresponding frequency if the deviation is of order $1s^{-1}$.
%about the location of the particle.
Thus, observing the values $\tilde{\omega}_1$ in Fig.~\ref{fig:BA}b, we estimate that in the experiment by BA $\omega_1$ is recovered for $m\approx80$, i.e.~after at least about 80 detection time steps have been taken into account.
The same analysis based on Fig. \ref{fig:ND1}b yields that for the particle localized at $L_1$ roughly $m\approx200$ time steps are  required for a reliable estimation of $\omega_1$.
%These, however, are not the relevant numbers.
This corresponds to a huge difference in the number of postselected photons $N_{\rm post}$ needed in the two cases
%We need to count the number of postselected photons necessary for the estimation of of $\omega_1$. There is a huge difference in the number of postselected photons $N_{\rm post}$ needed in the two cases:
\begin{align}\label{Npost}
    N_{\rm post}^{BA}=\sum_{k=1}^{80} N_k^{BA} \approx 22000, \quad \quad N_{\rm post}^{L_1}=\sum_{k=1}^{200} N_k^{L_1} \approx 25.
\end{align}
To obtain similar information about $L_1$ in the setup of BA the required number of photons was larger by a factor of about 1000 relative to the case where the photons actually were localized at $L_1$.
%To ensure that our results are not accidental we repeated the simulation and the analysis several times and observed similar behaviour.

\begin{figure}
\centering
\subfloat{
    \includegraphics[width=\textwidth]{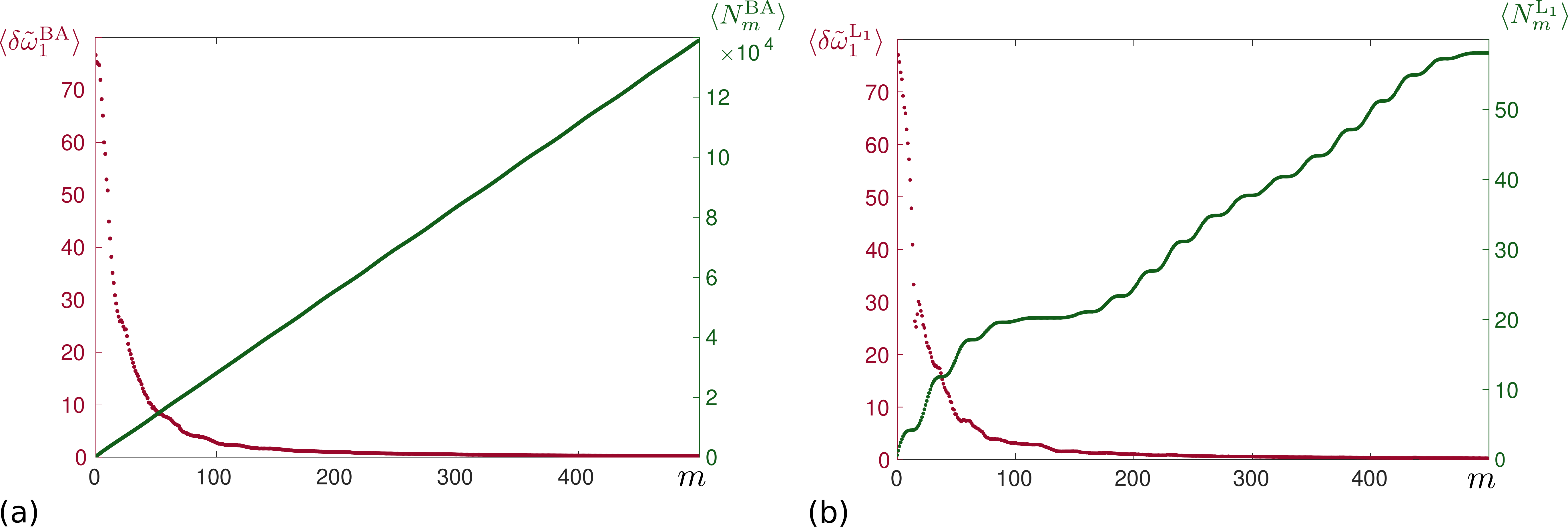}}
\caption{ {\bf Average deviation in estimation of $\omega_1$  in 500 runs of simulations.} %$N_m\equiv\sum_{k=1}^m N_k$ 
(a) Estimation of $\omega_1$ in BA experiment. (b) Estimation of $\omega_1$ in BA setup, conditioned on the nondemolition  detection of the photon at $L_1$.
$\langle N^\mathrm{BA}_m\rangle$ and $\langle N^\mathrm{L_1}_m\rangle$ are the average of total number of postselected photons after $m$ steps.
Note the difference in the scales for $\langle N^\mathrm{BA}_m\rangle$ and $\langle N^\mathrm{L_1}_m\rangle$.}
\label{fig:errN1}
\end{figure}

To ensure that our results are not accidental we repeated the simulation 500 times and plotted the average of the deviation  in the estimation $\delta  \tilde{\omega} \equiv |\tilde{\omega} -\omega|$ and the average number of postselected photons as a function of $m$ for the two cases, see Fig.~\ref{fig:errN1}.
We see from the graphs that in order to get the same precision of estimation in the BA method and in the case of photons present at $L_1$, the ratio of the numbers of postselected photons is even larger than 1000.

The experimental simulation shows that the  amount of information about $L_1$ obtained by each particle in the BA experiment is much less than the information that would be obtained by a particle which definitely was at $L_1$.
%{\color{red}Thus, we are not able to conclude that the likelihood of the photon having been at $L_1$ is similar to it having been at $L_2$, $L_3$, or $L_4$.}
%Thus, we are not forced to say that the particle in the BA experiment must have been at $L_1$.
In fact, the criterion of gained information tells us that the presence of the photons at $L_1$ is even smaller than the presence given by the weak value criterion according to which the presence of the photons in the BA setup is of the order $({\rm \bf P}_{L_1})_w\approx \epsilon =10^{-2}$.
%The alleged contradiction with statement S-C does not exist, since proposition i) is not strictly true and thus a properly understood S-C does not indicate {\color{red} a strong likelihood that the photon had been at $L_1$}.

The BA experiment does not show correctly the weak trace of the particle at $L_1$ because the information recorded on the photon of the ``leaked'' channel (in the terminology of BA), does not reach the detector undisturbed due to the presence of the phase shifter $\eta$.
It does properly show the photon presence at $L_2$, $L_3$, and $L_4$ (and its absence at $L_5$) because there is only one (two-mode) path from every one of these locations and even if the phase is shifted, the signal is not distorted.
As we will show below for the example of $L_4$, the information gain about these regions is not especially sensitive and of the same order as the information gain from the photon localized in these regions.

\begin{figure}
\centering
\subfloat{
    \includegraphics[width=\textwidth]{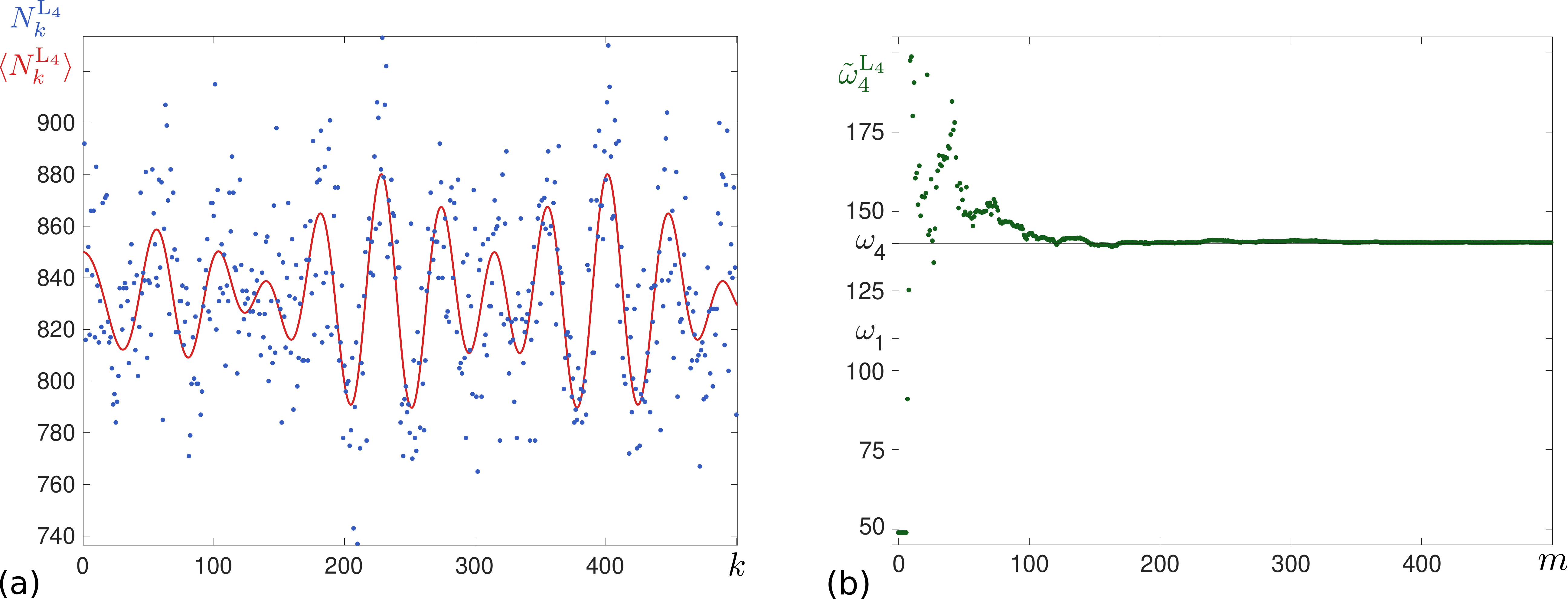}}
\caption{ {\bf Estimation of the frequency of the modulation at $L_4$ based on photons passing through $L_4$.}
(a) The points provide the numbers of detected photons $N_k^{L_4}$ in every time window $k$ conditioned on their localization at $L_4$ generated by a computer simulation  of sending $N_s$ photons with detection probability (\ref{P_kL_4}). 
The continuous line is the theoretical expectation value $\langle{N}_k^{L_4}\rangle$.
(b) The results of the estimation algorithm for the frequency $\omega_4$ (with all other parameters known).}
\label{fig:ND4}
\end{figure}

%2.6 Simulation to show NO hyper sensitivity in other parts of the experiment
%To show the difference in information gain between $L_1$ and $L_4$, 

\subsection{Abscence of hyper-sensitivity at location $L_4$}
\label{ssec::hyperSensL4}

We analyze the signal for the presence of the photon at $L_4$ in the BA experiment using the same dataset of postselected particles in Fig.~\ref{fig:BA}a. Performing the same estimation procedure as before, but treating all $\omega_i$ for $i\neq 4$ as known, yields the plot for $\Tilde{\omega}_4^{BA}(m)$ shown in Fig. \ref{fig:BA}b (green points). 

For comparison, adding a nondemolition measurement of the presence of the particle at $L_4$ yields the conditional probability of detection in $D$ 
\begin{align}\label{P_kL_4}
    P_k^{L_4} &\approx \frac{1}{6}[1+2\epsilon (\cos \omega_1t+\cos \omega_4t-\cos \omega_5t)].
\end{align}
%with $\bar{N}_k^{L_4} = N_s P_k^{L_4}$.
Contrary to the case of photons passing through $L_1$, %which yields the probability (\ref{non_L1}) quadratic in $\epsilon$, 
a large constant term similar to that in (\ref{BA}) remains present here. Thus, an amplified signal sensitivity, as seen for $L_1$, is not expected for $L_4$.
The generated dataset $ \{N_1^{L_4}, N_2^{L_4},...,N_m^{L_4}\}$, conditioned on full localization at $L_4$, is shown in Fig. \ref{fig:ND4}a, with
%The generated dataset $ \{N_1^4, N_2^4,...,N_m^4\}$, conditioned on full localization at $L_4$, is shown in Fig. \ref{fig:ND4}a, with
estimated values $\tilde{\omega}_4^{L_4}(m)$ shown in Fig. \ref{fig:ND4}b. 
We estimate from the plot that the fit result for $\omega_4$ starts to converge to the actual value starting from $m\approx175$.
(It is not surprising that we need a longer run than for estimation of $\omega_1$ because of the factor of 2 between the $\cos\omega_1 t$ term and the $\cos\omega_4 t$ term in (\ref{BA})).
In the simulation with photons localized at $L_4$ we get a good estimate of $\omega_4$ starting from $m \approx 100$.
For the estimation of $\omega_4$ there is no order of magnitude difference in the number of postselected photons $N_{\rm post}$ needed in the two cases
\begin{align}\label{Npost2}
    N_{\rm post}^{BA}=\sum_{k=1}^{175} N_k^{BA} \approx 50000, \quad \quad N_{\rm post}^{L_4}=\sum_{k=1}^{100} N_k^{L_4} \approx 80000.
\end{align}

\begin{figure}
\centering
\subfloat{
    \includegraphics[width=\textwidth]{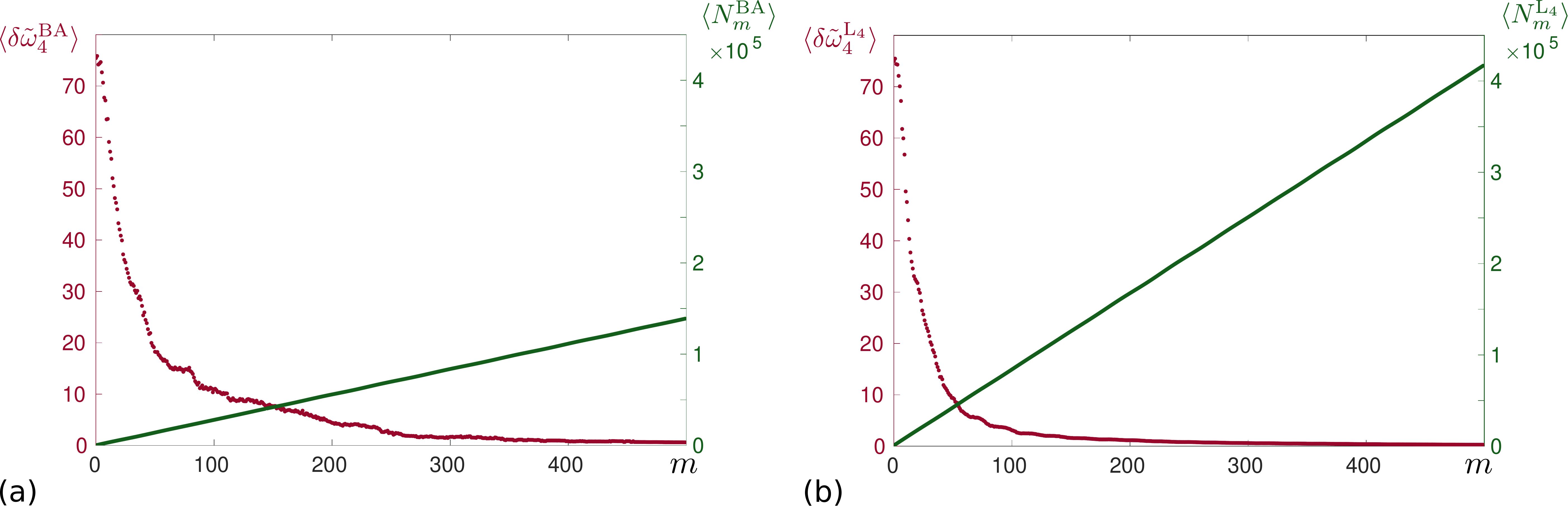}}
\caption{ {\bf Average deviation in  estimation of $\omega_4$ in 500 runs of simulations.} (a) Estimation of $\omega_4$ in BA experiment. (b) Estimation of $\omega_4$ in BA setup, conditioned on nondemolition  detecting of the photon at $L_4$. $\langle N^\mathrm{BA}_m\rangle$ and $\langle N^\mathrm{L_1}_m\rangle$ are the average of total number of postselected photons after $m$ steps.}
\label{fig:errN4}
\end{figure}

We repeated the simulation 500 times also for these two cases and obtained the average of the deviation in the estimation and the average of the number of the postselected photons $N_m\equiv\sum_{k=1}^m N_k$ as function of $m$, see Fig. \ref{fig:errN4}.
The results confirm that we need the same order of magnitude of postselected photons to obtain the same precision of the frequency estimation in the BA experiment relative to the case when the photons were known to be at $L_4$ by non-demolition measurement.
Therefore, we {\it can} consider the BA procedure as providing a signal that indicates the presence of the  photons  at $L_4$.
Similar conclusions can be made about $L_2$ and $L_3$. 

%2.7 Difference between amplification of signal and amplification of presence

\section{Simplified Bhati-Arvind experiment}
\label{sec::simplified}
%At first glance it might seem strange that similar experimental elements under some circumstances lead to the amplification of sensitivity and in others to amplification of presence, but this is a natural complication fully related to the fact that here, degrees of freedom of the pre- and postselected particle are used as the pointer and in effect only a partial postselection of the particle is performed to allow the application of S-C.
%No such considerations are necessary in the defining scenario of the WTA, when a local environment measures the trace and S-B is applied.

%--------------------------------------------------------------------------
%CONCLUSION
 
In our analysis above, following BA, we have considered their interferometric setup with multiple transmission modulators acting simultaneously in several locations.
%Multiple modulators in this six-port interferometer led to unnecessary complication.  
This led to a superposition of the various oscillating signals which made it non-trivial to explicitly calculate the information content with respect to particular local parameters $\omega_i$.
For a more clear illustration of the effect in question we now construct a simple modification of the mode of operation of the BA setup, which keeps the core of their method intact but provides a more direct way to quantify the amount of information about the disturbance carried by the postselected particles. 
%but makes the  analysis of its strengths and weaknesses more transparent. 
%allows to directly quantify the information content about each parameter in the signal, by explicitly calculating the error of trying to estimate the parameter from the signal.
This will demonstrate more clearly our claims about the increased sensitivity to a disturbance at $L_1$ and the role of the phase shifter $\eta$ in the BA setup.

The criterion of BA for presence at a particular location is that the postselected photons carry information about the disturbance at this location.
The disturbances are simultaneous modulations of transparency of beamsplitters at various locations with different frequencies.
In our simplified version of the BA proposal we replace one experiment with modulations of all beamsplitters by several experiments in everyone of which only one beamsplitter is operating while all others are replaced by mirrors.
This change allows also another simplification. %Instead of the periodic modulation we consider just two time intervals.
%
%We replace different  harmonic modulations of the operating beamsplitter  by a constant transmitivity $\epsilon$ during the first or the last half of the time of the experiment keeping full reflectivity during the other half.
%Instead of the periodic modulation we consider just two time intervals, in one of which the constant transmittivity $\epsilon$ in the particular beamsplitter is switched on, while in the other interval it is fully reflective.
%%While in the BA experiment the photons carry information about the frequencies of local disturbances, i
%%In the simplified version the photons carry information about a particular location: 
Instead of the periodic modulation we consider just two time windows, in one of which the constant transmittivity $\epsilon$ at a particular beamsplitter is switched on, while in the other time window the beamsplitter is fully reflective.
%The photons carry information about which time window (out of two) has the disturbance.
The photons carry information in which time window (out of two) the disturbance occurred.

In the experiment, we chose a location $L_i$ and randomly introduce a constant disturbance $\epsilon$, i.e.~add a constant transmittivity, in the first or second time window.
We propose the following strategy to extract information about the timing of the disturbance, from the distribution of detected photons between the two time windows, where $P_\epsilon$ and $P_0$ are the probabilities of postselection with and without disturbance respectively.
%We expect that the distribution of detected photons between the two windows will be correlated with the probabilities of detection with disturbance, $P_\epsilon$, and without disturbance, $P_0$.
If $P_\epsilon > P_0$, we infer that the disturbance was in the time window with the larger number of postselected photons.
Conversely, if $P_\epsilon < P_0$, we infer that the  disturbance was in the time window with the smaller number of postselected photons.
In the case that the numbers of photons in the two time windows are equal we refrain from any choice. %and count that case as an error!
%In the case that the numbers of photons in the two time windows are equal we can only choose randomly in which to assume the disturbance.
Let $P$ be the bigger probability out of $P_\epsilon$, $P_0$, and $p$ the smaller probability.
Then the probability of an error, i.e.~the probability of the failure to provide the correct inference of the window with disturbance, is the probability that we will get smaller or equal than half of the postselected photons in the window with higher probability of postselection
\begin{align}
\label{eq::proberror}
\mathrm{prob}(\mathrm{error}) = \sum_{N_P=0}^{\lfloor N/2 \rfloor}\binom{N}{N_P}
\Big(\frac{P}{P+p}\Big)^{N_P}
\Big(\frac{p}{P+p}\Big)^{N-N_P},
\end{align}
where $N_P$ is the number of postselected photons in the time window with probability of postselection $P$ and the sum goes over integers from $0$ to floor $N/2$.
Now we can apply our definition (\ref{infodefi}) where the local information is whether the disturbance happens in the first or the second time window.
We choose $1\%$ as the predefined error threshold of the probability of a correct inference which corresponds to the precision of the obtained information.
%We measure the precision of the information via the probability of a correct inference and choose $1\%$ as the predefined error threshold.
%If we define an error threshold of $1\%$
%If we allow 
%This allows to apply our definition (\ref{infodefi})
%\cNew{This probability of error allows to evaluate the amount of information per photon using definition (\ref{infodefi}).}
%The probability of detection with and without disturbance together with our postulate of an error less than $1\%$ in the declaration allows to estimate the amount of information carried by the photons according to our definition (\ref{infodefi}).

%\begin{align}
%\sum_{N_1=0}^{N}prob(N_1|N) H\bigl(N_1>N-N_1\bigr)> 0.99,
%\end{align}
%where $N_1$ is number of postselected photons in window with disturbance $\epsilon$. The probability of random preselected photon to be successfully postselected in the window with disturbance $\epsilon$ is $\tfrac{1}{2} P_\epsilon$. Now the probability of a single particle to be in window with the disturbance given that it was postselected is: 
%\begin{align}
%P_1=\frac{\tfrac{1}{2}P_\epsilon}{\tfrac{1}{2}P_\epsilon+\tfrac{1}{2}P_0}.
%\end{align}
%Thus the probability of postselecting $N_1$ particles in window with disturnace given $N$ postselected particles in total is:
%\begin{align}
%prob(N_1|N)=\binom{N}{N_1}P_1^{N_1}P_2^{N-N_1}.
%\end{align}

%%The BA method of getting this information is by observation of the effect of the disturbance on the probability of the postselection. 
In our simplified version of the BA experiment in the window without disturbance the probability of detection is $P^{BA}_0=\frac{1}{18}$.
%A transmission amplitude $\epsilon$ introduced at a beamsplitter  $L_i$ (keeping all other  $L_j$ to be mirrors) leads to the following probabilities of detection
In the time window with the disturbance at one of the beamsplitters $L_i$ the detection probability is given by the following expressions
\begin{align}\label{onebyone}
    P^{BA}_{L_1}\cong  \frac{1}{18}(1+4\epsilon), ~~ P^{BA}_{L_2}\cong  \frac{1}{18}(1-2\epsilon), ~~ P^{BA}_{L_3}\cong  \frac{1}{18}(1+2\epsilon), ~~ P^{BA}_{L_4}\cong  \frac{1}{18}(1+2\epsilon), ~~
    P^{BA}_{L_5}=  \frac{1}{18}.
\end{align}
These relations, replacing (\ref{BA}) of the original BA method, allegedly show the presence of the particle near $L_1$, $L_2$, $L_3$, and $L_4$ and absence at $L_5$.
%by changing the probability of detection in the same order, and absence at $L_5$ because the disturbance there does not affect the probability of detection. 
The signal of the first order of $\epsilon$ is present in all cases except of $L_5$.
%Now, accepting the definition of information (\ref{infodefi}) of the photon corresponding to less that $1\%$ error and taking constant disturbance $ \epsilon=10^{-2}$ in every specified location  of the BA experiment, we get the following amount of information carried by photons in experiments with disturbances in various locations
Now, applying (\ref{infodefi}) with (\ref{eq::proberror}) and assuming a constant disturbance $\epsilon=10^{-2}$, we calculate the amount of information $\mathcal{I}^{BA}_{L_1}$ carried by photons with disturbances at various locations as shown in the first row of Table \ref{table::information}.
%\begin{align}\label{info}
%    \mathcal{I}_{L_1}= 7.06 \cdot 10^{-5},~~~~~
%    \mathcal{I}_{L_2}=1.88\cdot 10^{-5},~~~~~
%    \mathcal{I}_{L_3}=1.81\cdot 10^{-5},~~~~~
%    \mathcal{I}_{L_4}=1.81\cdot 10^{-5},~~~~~
%    \mathcal{I}_{L_5}= 0.
%\end{align}

Although we introduce a transmission amplitude of the same value $\epsilon$ at every beamsplitter $L_i$, as already argued above it would be a mistake to assume that in general the amount of information the postselected photons carry about local disturbances $\mathcal{I}^{BA}_{L_i}$ faithfully characterizes the presence of the particle at the corresponding locations.
We have to take into account a possibly different sensitivity of the probability of the detection on the transmission amplitude  $\epsilon$ introduced at different locations $L_i$.
Thus, as already argued in section \ref{sec::theo::criteria}, the proper measure of the presence of a particle at a particular location $\mathcal{M}_{L_i}$ is the ratio of information content of the photons obtained in this experiment $\mathcal{I}^{BA}_{L_i}$ and the information content of photons which actually passed this location $\mathcal{I}^{L_i}_{L_i}$ (the superscript describes where photon comes from, the subscript where the disturbance takes place)
\begin{align}
\label{eq:measurePr}
    \mathcal{M}^{BA}_{L_i} = \mathcal{I}^{BA}_{L_i} / \mathcal{I}^{L_i}_{L_i}.
    %\mathcal{M}^{BA}_{L_i}=\frac{\mathcal{I}^{BA}_{L_i}}{\mathcal{I}^{L_i}_{L_i}}.
\end{align}

%To test this dependence
%To characterize the amount of information $\mathcal{I}^{L_i}_{L_i}$ of photons passed through certain location $L_i$, we consider the original experiment with additional non-demolition measurement, which finds the particle at this location. To this end we compare two cases of the probability of the postselection detection of the photons passing through each location $L_i$: 
%$P^{L_i}_0$, when it works as a mirror, and  $P^{L_i}_{L_i}$, when we introduce transmission amplitude $\epsilon$ at $L_i$ . 

As in Section \ref{sec::fit} we consider localized photons by additionally conditioning on successful non-demolition measurements at the respective locations $L_i$.
Given that the photon was found at $L_i$, when there is no disturbance (all devices $L_i$ reflect $100\%$), the conditional probabilities of detection by detector $D$ are
%, given that the particle was found at $L_i$, are:
\begin{align}\label{onebyoneLi0}
    P^{L_1}_0=0  ,~~~~~ P^{L_2}_0=  \frac{1}{6},~~~~~ P^{L_3}_0=  \frac{1}{6},~~~~~ P^{L_4}_0=  \frac{1}{6},~~~~~ P^{L_5}_0=  \frac{1}{3}.
\end{align}
%We use the concept of conditional probability of passing through $L_i$ in BA setup for defining properly the photon state in the two channels of each $L_i$. 
Note that contrary to the original setup of BA, in our simplified version without disturbances at other locations the photons cannot be found at $L_5$, so the meaning of ``the photon found at $L_5$'' is instead that the photon was placed in the main (upper) channel of $L_5$.
 
%Now consider photons conditioned to be at a particular $L_i$ with transmission amplitude $\epsilon$  while all other devices $L_j$ are mirrors.
The probabilities of detection of photons conditioned to be at a particular $L_i$ with transmission amplitude $\epsilon$ at the same location are
\begin{align}\label{onebyoneLi}
    P^{L_1}_{L_1}\cong \frac{1}{3}\epsilon^2, \quad 
    P^{L_2}_{L_2}\cong \frac{1}{6}(1-2\epsilon), \quad
    P^{L_3}_{L_3}\cong \frac{1}{6}(1-2\epsilon), \quad P^{L_4}_{L_4}\cong \frac{1}{6}(1+2\epsilon), \quad P^{L_5}_{L_5}\cong \frac{1}{3}(1-2\epsilon).
\end{align}
We see that local disturbances $\epsilon$ introduced at locations  $L_2$, $L_3$, $L_4$, $L_5$ cause identical changes in the probability of postselection while the same local disturbance at $L_1$ has a very different effect. % causes a very different change of the probability of postselection. 
The amount of information carried by postselected photons localized at the different locations characterizes their sensitivity to disturbances and is shown in the second row of Table \ref{table::information}.
%\begin{align}\label{infoseparate}
%    \mathcal{I}^{L_1}_{L_1}=1,~~~~~
%    \mathcal{I}^{L_2}_{L_2}=1.81\cdot 10^{-5},~~~~~
%    \mathcal{I}^{L_3}_{L_3}=1.81\cdot 10^{-5},~~~~~
 %   \mathcal{I}^{L_4}_{L_4}=1.81\cdot 10^{-5},~~~~~
 %   \mathcal{I}^{L_5}_{L_5}=1.81\cdot 10^{-5} ,
%\end{align}
Now, we can use (\ref{eq:measurePr}) to provide a quantitative characterisation of the magnitude of the presence of the photons at various locations, shown in the third row of Table \ref{table::information}.
%The measure of presence $\mathcal{M}^{BA}_{L_i}$ is shown in the third row of Table \ref{table::information}.
%the ratio of the information about the location $L_i$ carried by the photon in the BA experiment $\mathcal{I}^{BA}_{L_i}$
%(\ref{info})
%and the sensitivity $\mathcal{I}^{L_i}_{L_i}$,
%(\ref{infoseparate}) 
%provides a quantitative characterisation of the magnitude of the presence of the photons at various locations. 
%\begin{align}\label{inforatio}
%   \= \frac{\mathcal{I}^{BA}_{L_1}}{\mathcal{I}^{L_1}_{L_1}}=7.06 \cdot 10^{-5},~~~~~
%    \frac{\mathcal{I}^{BA}_{L_2}}{\mathcal{I}^{L_2}_{L_2}}=1.04,~~~~~
%    \frac{\mathcal{I}^{BA}_{L_3}}{\mathcal{I}^{L_3}_{L_3}}=1,~~~~~ 
%    \frac{\mathcal{I}^{BA}_{L_4}}{\mathcal{I}^{L_4}_{L_4}}=1,~~~~~
%    \frac{\mathcal{I}^{BA}_{L_5}}{\mathcal{I}^{L_5}_{L_5}}= 0.
%\end{align}

These results qualitatively agree with the WTA, e.g.~as calculated in \cite{Bhati}. % of projection on the locations in the BA experiment as shown in the fourth row of Table \ref{table::information}.
%\begin{align}\label{weakpresence}
%    ({\mathbf{P}}_{L_1})_w=0,~~~~~
%    ({\mathbf{P}}_{L_2})_w=1,~~~~~
%    ({\mathbf{P}}_{L_3})_w=-1,~~~~~
%    ({\mathbf{P}}_{L_4})_w=1,~~~~~
%    ({\mathbf{P}}_{L_5})_w= 0.
%\end{align}
The presence characterized by the relative amount of information about a local disturbance $\mathcal{M}^\mathrm{BA}_{L_i}$, describes, of course, only the magnitude of the presence and not the modification of the effective weak local interactions specified by (in general) complex numbers representing the weak values \cite{PNAS}.
What is shown clearly in this simplified version of the BA experiment is the anomalous sensitivity for a disturbance at $L_1$ which explains how the photons are ``lying'' about their presence there.

\renewcommand{\tScale}{1.3}
\begin{table}[h]
%\raisebox{distance}[extend-above][extend-below]{text}
%\begin{tabular}{|c||c|c|c|c|c|}
\begin{tabular}{|Sc||Sc|Sc|Sc|Sc|Sc|}
\hline
 & \scalebox{\tScale}{$L_1$} & \scalebox{\tScale}{$L_2$} & \scalebox{\tScale}{$L_3$} & \scalebox{\tScale}{$L_4$} & \scalebox{\tScale}{$L_5$} \\
\hline
\hline
\scalebox{\tScale}{$\mathcal{I}^{BA}_{L_i}$} & $7.06 \cdot 10^{-5}$ & $1.88 \cdot 10^{-5}$ & $1.81 \cdot 10^{-5}$ & $1.81 \cdot 10^{-5}$ & $0$ \\
\hline
\scalebox{\tScale}{$\mathcal{I}^{L_i}_{L_i}$} & $1$ & $1.88 \cdot 10^{-5}$ & $1.88 \cdot 10^{-5}$ & $1.81 \cdot 10^{-5}$ & $1.88 \cdot 10^{-5}$ \\
\hline
\scalebox{\tScale}{$\mathcal{M}^{BA}_{L_i}$} & $7.06 \cdot 10^{-5}$ & $1$ & $0.96$ & $1$ & $0$ \\
\hline
\scalebox{\tScale}{$\mathcal{I}^{BA'}_{L_i}$} & $0$ & $1.88 \cdot 10^{-5}$ & $1.81 \cdot 10^{-5}$ & $1.88 \cdot 10^{-5}$ & $0$ \\
\hline
\scalebox{\tScale}{$\mathcal{I}^{L_i'}_{L_i}$} & $1.88 \cdot 10^{-5}$ & $1.88 \cdot 10^{-5}$ & $1.88 \cdot 10^{-5}$ & $1.88 \cdot 10^{-5}$ & $1.88 \cdot 10^{-5}$ \\
\hline
\scalebox{\tScale}{$\mathcal{M}^{BA'}_{L_i}$} & $0$ & $1$ & $0.96$ & $1$ & $0$ \\
\hline
\end{tabular}
%\label{table::information}
\captionlistentry{}
\begin{flushleft}
\textbf{Table 1: Amount of carried information and measure of presence.} $\mathcal{I}^{BA}_{L_i}$ represent the amount of information carried by the postselected photon in the BA experiment, $\mathcal{I}^{L_i}_{L_i}$ the amount for photons coming from location $L_i$. $\mathcal{M}^{BA}_{L_i}$ is the measure of presence of a photon in the BA experiment.
$\mathcal{I}^{BA'}_{L_i}$, $\mathcal{I}^{L_i'}_{L_i}$ and $\mathcal{M}^{BA'}_{L_i}$ stand for the same quantities in the BA experiment without the phase shifter $\eta$.
%$\mathcal{I}=1/N_{min}$ calculated with disturbance $\epsilon=10^{-2}$ and threshold probability of error $\alpha = 1\%$. The measure if presence $\mathcal{M}^{BA}_{L_i}=\mathcal{I}^{BA}_{L_i}/\mathcal{I}^{L_i}_{L_i}$ qualitatively agrees with the weak value of the projection $(\mathbf{P}_{L_i})_w$ on a certain location $L_i$.
\end{flushleft}
\label{table::information}
\end{table}

%\setlength{\tabcolsep}{14pt} % Default value: 6pt
%\renewcommand{\arraystretch}{1.5} % Default value: 1
%\begin{table}[ht]
%\begin{tabular}{|llll|}
%\hline
%\multicolumn{1}{|l|}{}      & \scalebox{1.3}{$\mathcal{I}^{BA}$}   & \scalebox{1.3}{$\mathcal{I}^{BA'}$}  & \scalebox{1.3}{$\mathcal{I}^{L_i}$}  \\ 
%\hline
%\multicolumn{1}{|l|}{\scalebox{1.3}{$L_1$}} & $7.06 \cdot 10^{-5}$ & $0$                  & $1$                  \\
%\multicolumn{1}{|l|}{\scalebox{1.3}{$L_2$}} & $1.88 \cdot 10^{-5}$ & $1.88 \cdot 10^{-5}$ & $1.81 \cdot 10^{-5}$ \\
%\multicolumn{1}{|l|}{\scalebox{1.3}{$L_3$}} & $1.81 \cdot 10^{-5}$ & $1.81 \cdot 10^{-5}$ & $1.81 \cdot 10^{-5}$ \\
%\multicolumn{1}{|l|}{\scalebox{1.3}{$L_4$}} & $1.81 \cdot 10^{-5}$ & $1.88 \cdot 10^{-5}$ & $1.81 \cdot 10^{-5}$ \\
%\multicolumn{1}{|l|}{\scalebox{1.3}{$L_5$}} & $0$                  & $0$                  & $1.81 \cdot 10^{-5}$ \\
%\hline

%\end{tabular}
%\begin{flushleft}
%{\bf Table 1: Information per Photon.} $\mathcal{I}=1/N_{min}$ calculated with disturbance $\epsilon=10^{-2}$ and threshold probability of error $\alpha = 1\%$.
%\end{flushleft}
%\end{table}

To reveal the source of increased sensitivity it is instructive to consider the amount of information carried by the pre- and postselected photons in our simplified experiment when the $\pi$ phase shifter $\eta$ is removed, signified as $BA^\prime$ (In the original setup of BA, removing the phase shifter removes the term $\frac{2}{9} \epsilon \cos \omega_1t$ from (\ref{BA})).
%It is instructive to consider the information content of pre- and postselected photons in the BA interferometer without the $\pi$ phase shifter $\eta$, which we will signify as $BA'$ (in the original BA method, removing  the phase shifter removes the term $\frac{2}{9} \epsilon  \cos \omega_1t$ from (\ref{BA})).
In this case the probabilities are given by
\begin{align}\label{onebyoneBA'}
    P^{BA'}_{L_1} = \frac{1}{18}, \quad
    P^{BA'}_{L_2} \cong \frac{1}{18}(1-2\epsilon), \quad
    P^{BA'}_{L_3} \cong \frac{1}{18}(1+2\epsilon), \quad
    P^{BA'}_{L_4} \cong \frac{1}{18}(1-2\epsilon), \quad
    P^{BA'}_{L_5} = \frac{1}{18},
\end{align}
while the probability without disturbance stays $P_0^{BA'} = \frac{1}{18}$.
The corresponding amount of information $\mathcal{I}^{BA'}_{L_i}$ is shown in the fourth row of Table \ref{table::information}.
Note, that the amount of information 0 in the case of $\mathcal{I}^{L_1}_{L_1}$ comes from the fact that in this case the probability of error (\ref{eq::proberror}) is never less than $1\%$ and thus no $N_\mathrm{min}$ exists for which criterion (\ref{infodefi}) is satisfied.
%\begin{align}\label{infoDanan}
%    \mathcal{I}^{BA'}_{L_1}=0,~~~~~
%    \mathcal{I}^{BA'}_{L_2}=1.88\cdot 10^{-5},~~~~~
%    \mathcal{I}^{BA'}_{L_3}=1.81\cdot 10^{-5},~~~~~
%    \mathcal{I}^{BA'}_{L_4}=1.88\cdot 10^{-5},~~~~~
%    \mathcal{I}^{BA'}_{L_5}= 0.
%\end{align}

Again, to obtain a faithful measure of presence, we need to compare this to the amount of information provided by photons which had actually been at these locations. 
There is a difficulty here: photons located at $L_1$ have zero probability to be postselected, such that no $\mathcal{I}^{L_1'}_{L_1}$ can be calculated.
%To overcome this difficulty, we can introduce a tiny imbalance of $\delta$ in the inner interferometer. All the other calculations presented are unaffected by this procedure in the limit of $\delta \to 0$ wherefore $\delta$ can be set to $0$.
To overcome this difficulty, we can introduce a tiny imbalance $\delta$ in the inner interferometer.
%and set $\lim_{\delta \to 0}$ in the further process. 
Assuming $\delta$ much smaller than $\epsilon$ and all other parameters of the problem, the effect of the imbalance $\delta$ is significant only when the probability of postselection has been zero before.
%Assuming $\delta$ much smaller than $\epsilon$ and all other parameters of the problem, the amount of information per photon is affected significantly only where the probability of postselection has been zero before.
The probabilities of detection of photons conditioned to be at a particular $L_i$ with transmission amplitude $\epsilon$ and imbalance $\delta$ are
\begin{align}\label{onebyoneLi'}
    P^{L_1'}_{L_1}\cong  \frac{1}{3}\delta(1-2\epsilon),~~ P^{L_2'}_{L_2}\cong  \frac{1}{6}(1-2\epsilon),~~ P^{L_3'}_{L_3}\cong  \frac{1}{6}(1-2\epsilon),~~ P^{L_4'}_{L_4}\cong  \frac{1}{6}(1-2\epsilon),~~ P^{L_5'}_{L_5}\cong  \frac{1}{3}(1-2\epsilon).
\end{align}
In the time window with no disturbance the probabilities are 
\begin{align}\label{onebyoneLi0'}
    P^{L_1'}_0=\frac{\delta}{3}  ,~~~~~ P^{L_2'}_0=  \frac{1}{6},~~~~~ P^{L_3'}_0=  \frac{1}{6},~~~~~ P^{L_4'}_0=  \frac{1}{6},~~~~~ P^{L_5'}_0=  \frac{1}{3}.
\end{align}
%The resulting amount of information $\mathcal{I}^{L_i}_{L_i}$, presented in the fifth row of Table \ref{table::information}, together with $\mathcal{I}^{BA'}_{L_i}$, leads us to the quantitative measure of presence $\mathcal{M}^{BA'}_{L_i}$ as shown in the sixth row of Table \ref{table::information}. 
The resulting amount of information $\mathcal{I}^{L_i'}_{L_i}$, with $\epsilon =10^{-2}$ and $\delta\to 0$, is presented in the fifth row of Table \ref{table::information}. 
%It remains the same 
Since it was obtained in the limit $\delta \to 0$, it can be considered together with $\mathcal{I}^{BA'}_{L_1}$, which was calculated for $\delta = 0$.
(Also for the calculation of $\mathcal{I}^{BA'}_{L_1}$ the same approach with a small $\delta$ is possible and yields the same results.)
Thus, we can calculate the quantitative measures of presence $\mathcal{M}^{BA'}_{L_i}$ at $L_i$ as shown in the sixth row of Table \ref{table::information}. 

The BA experiment without the phase shifter is equivalent to the original nested interferometer experiment \cite{Danan}. 
The experiment correctly shows the presence of the particle near  $L_2$, $L_3$, and $L_4$ ($\mathcal{M}^{BA'}_{L_{i}}\approx 1 $) and no presence at $L_1$ and $L_5$ ($\mathcal{M}^{BA'}_{L_{i}} = 0$).
%This information can be read off from the quantitative measure of presence $\mathcal{M}^{BA'}_{L_i}$ as well as from the signal in the experiment proportional to $\mathcal{I}^{BA'}_{L_i}$.
In this case the same information also can be read off from the signal which is proportional to $\mathcal{I}^{BA'}_{L_i}$, so the photons are not ``lying'' in this experiment.

\section{Conclusions}
\label{sec::summary}

%At first glance it might seem strange that similar experimental elements under some circumstances lead to the amplification of sensitivity and in others to amplification of presence, but this is a natural complication fully related to the fact that here, degrees of freedom of the pre- and postselected particle are used as the pointer and in effect only a partial postselection of the particle is performed to allow the application of S-C.
%No such considerations are necessary in the defining scenario of the WTA, when a local environment measures the trace and S-B is applied.

%--------------------------------------------------------------------------
%CONCLUSION
BA by and large correctly describe the presence of photons inside their interferometer according to WTA, i.e.~proper versions of S-A and S-B.
The weak value of the presence at $L_1$ is of order $\epsilon$, while it is of order 1 at $L_2$, $L_3$, and $L_4$.
The apparent contradiction between the presence of information about the frequency of the modulation at $L_1$ of detected photons which, according to weak value criterion, have not been there, is resolved by analysing the quantitative details.
Vaidman's definition of the presence of a particle in a particular location is that the local trace it leaves is of the order of the trace a localised particle would leave there.
The definition of the presence according to the information gain about local properties (S-C) also must be quantitative. %: a single photon passing through the setup of BA tells us essentially nothing about the frequency of the modulation. %in $L_1$.
We have shown that the information gain about location $L_1$ of a photon which actually passed through $L_1$, i.e.~was detected by an additional non-demolition measurement in the BA setup, is much larger than the information gain of the photon in the original BA experiment, and thus we are not forced to conclude that the photons in the BA experiment have been at $L_1$.
The nonvanishing signal from the modulation at $L_1$ is explained by the nonvanishing presence there.
However, this presence is not of order 1 which would correspond to Vaidman's definition of presence \cite{past}, but is a ``secondary presence'' \cite{tracingthepast}.
%in $L_1$ \cite{past} and is named ``secondary presence'' \cite{tracingthepast}.

The photons of the BA experiment provide a similar amount of information about the frequencies of the modulations at $L_1$, $L_2$, $L_3$, and $L_4$.
This information tells us correctly that the photons were at $L_2$, $L_3$, and $L_4$.
However, the photons lie about their presence at $L_1$, which is negligibly small according to the WTA: the strong signal is explained by the interference effect leading to a large amplification of the sensitivity to a particular interaction taking place at $L_1$.
The weak values in the BA experiment represent faithfully the past of the particle as it is defined by Vaidman in \cite{past}.

In the BA method the measurable criterion for presence is the amount of information the postselected photons carry about some disturbance at a particular location. 
Obviously, the strength of the local disturbance is relevant but it is not the only factor. 
One has to consider the efficiency of the transfer of the information about the disturbance to the photons detected in the relevant output port.  
In the BA setup the disturbance created at $L_1$ is transferred to the postselected photons in an anomalously strong way. 
One way to explain this is that only the disturbed part is postselected because the undisturbed part of the photon state interferes destructively toward the output port. 
BA did not consider this in their analysis and made it possible for the postselected photons to lie about where they have been. 

%\section*{Declaration of competing interest}
%The authors declare that they have no known competing financial interests or personal relationships that could have appeared to influence the work reported in this paper.

\section*{Acknowledgements}

This work has been supported in part by the National Science Foundation Grant No. 1915015 and the U.S.-Israel Binational Science Foundation Grant No.~735/18 and by the Israel Science Foundation Grant No.~2064/19.
Furthermore, we want to acknowledge support by the DFG both under the project Beethoven 2 No.~WE2541/7-1 and under Germany's Excellence Strategy EXC-2111 390814868.

%unsrt
\bibliographystyle{unsrt}

%\bibliography{bib_com}

\begin{thebibliography}{10}

\bibitem{ABL}
Y. Aharonov, P. Bergmann,  and J. Lebowitz.
\newblock Time symmetry in the quantum process of measurement.
\newblock {\em Phys. Rev. }, 134, B1410, 1964.

\bibitem{AV90}
Y. Aharonov and L.~Vaidman.
\newblock Properties of a quantum system during the time interval between two
  measurements.
\newblock {\em Phys. Rev. A}, 41, 11, 1990.

\bibitem{AV91}
Y.~Aharonov and L.~Vaidman.
\newblock Complete description of a quantum system at a given time.
\newblock {\em J. Phys. A: Math. Gen.}, 24, 2315, 1991.



\bibitem{past}
L.~Vaidman.
\newblock Past of a quantum particle.
\newblock {\em Phys. Rev. A}, 87, 052104, 2013.





\bibitem{Bhati}
R.~S. Bhati and Arvind.
\newblock Do weak values capture the complete truth about the past of a quantum
  particle?
\newblock {\em Phys. Lett. A}, 429, 127955, 2022.


\bibitem{Danan}
A.~Danan, D.~Farfurnik, S.~Bar-Ad, and L.~Vaidman.
\newblock Asking photons where they have been.
\newblock {\em Phys. Rev. Lett.}, 111, 240402, 2013.

\bibitem{Dove}
M.~A. Alonso and A.~N. Jordan.
\newblock Can a dove prism change the past of a single photon?
\newblock {\em Quant. Stud.: Mat. Found.}, 2, 255, 2015.


\bibitem{KwiatSpinHall}
O.~Hosten and P.~Kwiat
\newblock Observation of the Spin Hall Effect of Light via Weak Measurements
\newblock {\em Science}, 319, 787, 2008.

\bibitem{Dixon}
P.B.~Dixon, D.J.~Starling, A.N.~Jordan, and J.C Howell,
\newblock Ultrasensitive beam deflection measurement via interferometric weak value amplification.
\newblock {\em Phys. Rev. Lett.}, 102, 173601, 2009.

 

\bibitem{PNAS}
J.~Dziewior, L.~Knips, D.~Farfurnik, K.~Senkalla, N.~Benshalom, J.~Efroni,
  J.~Meinecke, S.~Bar-Ad, H.~Weinfurter, and L.~Vaidman.
\newblock Universality of local weak interactions and its application for
  interferometric alignment.
\newblock {\em Proc. Natl. Acad. Sci. USA}, 116, 2881, 2019.

\bibitem{Groen}
J.~P.~Groen, D.~Rist\`{e}, L.~Tornberg, J.~Cramer, P.~C.~de~Groot, T.~Picot,
  G.~Johansson, and L.~DiCarlo.
\newblock Partial-measurement back-action and non-classical weak values in a superconducting circuit.
\newblock {\em Phys. Rev. Lett.}, 111, 090506, 2013.

\bibitem{Steinberg}
M.~Hallaji, A.~Feizpour, G.~Dmochowski, J.~Sinclair, and A.~M.~Steinberg .
\newblock Weak-value amplification of the nonlinear effect of a single photon
\newblock {\em Nat. Phys.}, 13, 540, 2017.


\bibitem{Lying}
L.~Vaidman and I.~Tsutsui.
\newblock When photons are lying about where they have been.
\newblock {\em Entropy}, 20, 538, 2018.


\bibitem{Rempe}
A. Reiserer, S. Ritter and G. Rempe,
\newblock Nondestructive detection of an optical photon.
\newblock {\em Science}, 342, 1349, 2013.




\bibitem{tracingthepast}
L.~Vaidman 
\newblock Tracing the past of a quantum particle.
\newblock {\em Phys. Rev. A}, 89, 024102, 2014.



%\bibitem{BoydAngRot}
%O.~S.~Magaña-Loaiza, M.~Mirhosseini, B.~Rodenburg, and R.~W.~Boyd
%\newblock Amplification of Angular Rotations Using Weak Measurements
%\newblock {\em Phys. Rev. Lett.}, 112, 200401, 2014.

\end{thebibliography}

\end{document}